\documentclass[a4,11pt]{article}

\usepackage{amssymb}
\usepackage{mathtools}
\usepackage{amsmath} 

\usepackage{xcolor}
\usepackage{tikz}

\usepackage{authblk}
\usepackage[numbers]{natbib}
\usepackage{siunitx}

\newcommand{\fig}{./figures/}

\begin{document}

\title{3D non-conforming mesh model for flow in fractured porous media using Lagrange multipliers}
\author[1]{Philipp~Sch\"{a}dle}
\author[2]{Patrick~Zulian}
\author[1]{Daniel~Vogler}
\author[1]{Sthavishtha~Bhopalam~R.}
\author[2]{Maria~G.~C.~Nestola}
\author[1]{Anozie~Ebigbo}
\author[2]{Rolf~Krause}
\author[1]{Martin~O.~Saar}
\affil[1]{Geothermal Energy and Geofluids Group, Department of Earth Sciences,~ETH~Z\"urich,~8092~Z\"urich,~Switzerland}
\affil[2]{Institute~of~Computational~Science,~USI~Lugano,~6904~Lugano,~Switzerland}

\date{Arxiv version from December 23, 2018 submission}

\maketitle
%%%
%%%
%%% ABSTRACT
%%%
%%%
\begin{abstract}
%%% OVERVIEW
This work presents a modeling approach for single-phase flow in 3D fractured porous media with non-conforming meshes. 
To this end, a Lagrange multiplier method is combined with a parallel $L^2$-projection variational transfer approach. 
This Lagrange multiplier method enables the use of non-conforming meshes and depicts the variable coupling between fracture and matrix domain.
The $L^2$-projection variational transfer allows general, accurate, and parallel projection of variables between non-conforming meshes (i.e.\ between fracture and matrix domain). \\
%%% WHAT WAS DONE
Comparisons of simulations with 2D benchmarks show good agreement, and the method is further validated on 3D fracture networks by comparing it to results from conforming mesh simulations which were used as a reference. 
Application to realistic fracture networks with hundreds of fractures is demonstrated. 
Mesh size and mesh convergence are investigated for benchmark cases and 3D fracture network applications. 
Results demonstrate that the Lagrange multiplier method, in combination with the $L^2$-projection method, is capable of modeling single-phase flow through realistic 3D fracture networks.
\end{abstract}
%
%%%
%%% KEYWORDS
%%%
\section{Introduction} \label{sec:introduction}
%
%%% GENERAL
Fractured rock formations in the subsurface are of crucial importance in a variety of reservoir applications, such as geothermal energy extraction, 
CO$_2$ sequestration, nuclear waste storage, and unconventional oil and gas recovery 
\cite{tester_2006,mcclure_2014b,bond_2003,bonnet_2001,rasmuson_1986,amann_2018}.
As fluid flow velocities in fractures are often magnitudes higher than in the rock matrix, individual fractures as well as fracture networks commonly govern the overall fluid transport characteristics of the entire fracture-dominated porous medium. 
Here, the geometric fracture configuration and the hydraulic properties of individual fractures, such as fracture permeability fields, largely determine, where preferential fluid flow may occur, as fractures with particularly high or low permeability can act as flow conduits or "bottlenecks", respectively \citep{dreuzy_2012,zimmerman_1991,ebigbo_2016}. Furthermore, \citet{ahkami_2018} describe and visualize experimentally that the permeability of the porous-medium matrix influences fluid flow in the fractures of a fractured porous medium.\\
\\
An in-depth understanding of processes in fractured rock masses thus requires knowledge about the hydraulic parameters of each fracture in a fracture network. 
As these parameters are notoriously difficult to obtain in the subsurface, reliance on stochastic investigations are often required, where hundreds or more system realizations typically have to be performed to assess uncertainties \cite{berkowitz_2002,cacas_1990,hobe_2018,neuman_2005}. 
To facilitate solving large numbers of numerical simulations of fluid flow through fracture-network systems in three dimensions (3D), highly efficient and accurate numerical methods and mesh generation approaches are required.
\\
Generally, in numerical models, two method classes are used when representing fractured porous media, i.e. fractures embedded in a porous-medium matrix.
Fractures are either represented by a continuum approach \cite{warren_1963,barenblatt_1960,kazemi_1969,kazemi_1976} or as discrete domains in a numerical mesh \cite{noorishad_1982,baca_1984}. 
In the continuum approach, the fractures and the porous-medium matrix share the same geometric mesh with separate continua. The respective flow properties are obtained by upscaling and information needs to be transfered between the continua. 
In contrast, the classic discrete-domain approach explicitly meshes both fractures and porous media, with the two meshes conforming at the boundaries of the domains (i.e.\ conforming numerical mesh).
Since fracture configurations in fracture networks can be arbitrarily complex, mesh generation for discrete fracture networks (DFN) \cite{hyman_2015,pichot_2010,pichot_2012,fumagalli_2019} or discrete fracture models (DFM) \cite{cacace_2015,holm_2006,blessent_2009} with the background matrix can be very difficult and time consuming. 
Due to the large length-to-width ratio of fractures and the need to overcome very small elements in the fracture domain, fractures may be represented by lower-dimensional elements, e.g. \cite{karimi_2003,bogdanov_2003,monteagudo_2004,helmig_1997}. 
Still, generating smooth matrix meshes on, and around, fracture intersections and fracture tips can lead to very small elements and significant increases in the number of degrees of freedom. 
In contrast, areas in the (porous) medium that are void of fractures might contain elements with large edge lengths, leading to
large differences in element size, compared to elements close to fractures.
The strong influence of fractures on fluid mass and energy transfer processes in a wide range of applications, as mentioned above, and the associated difficulties in model generation, have rendered related method improvements an area of active research. \\
\\
%%% NUMERICAL METHODS
The aforementioned shortcomings of classic discrete-domain approaches have led to an increased focus on the development of numerical methods that allow the use of independent meshes for the fracture domain and the matrix domain. 
Such non-conforming mesh approaches might be based on mortar methods \cite{frih_2012,boon_2018}, where the mesh for the discrete fractures and the matrix are required to align geometrically, but consist of independent discretizations. 
Larger flexibility is offered by methods that handle fracture and matrix meshes separately (i.e.\ no aligned geometries).
Such methods exist for finite volume schemes, e.g.\ (p)EDFM \cite{hajibeygi_2011,tene_2017,moinfar_2014}
and for XFEM-based approaches \cite[and references therein]{flemisch_2016} for finite elements.
A review of existing mathematical and conceptual models for flow in fractured porous media is given by \citet{berre_2018}.
Recently, \citet{koeppel_2018a} proposed a Lagrange multiplier method for a non-conforming finite element formulation. 
This method enables the use of independent meshes for fractures and matrices by applying variational transfer between the two mesh domains. 
The variational transfer allows projection of variables between the fractures and the matrix domain.
With the geometric mapping between the fractures and the matrix domain established, the Lagrange multiplier accounts for the variable coupling at the domain boundaries.
In contrast to XFEM methods, this approach does not enrich the finite element space locally, so that the pressure across the fractures is assumed to be continuous and the fractures have a higher permeability than the matrix.
\citet{koeppel_2018a} developed a general Lagrange multiplier method for 2D or 3D model domains. They further
show the uniqueness of the solution for the primal formulation of the continuous problem. 
Due to the large technical complexities in 3D, they focused on the implementation and verification in 2D.\\
\\
%%% OUR APPROACH
However, flow through fractured rock formations is governed by 3D effects, as strong heterogeneities affect flow properties in the fracture and porous-medium domains \citep{tsang_1998,dreuzy_2012,vogler_2018,vogler_2018b}.
To apply the Lagrange multiplier method to 3D fracture networks, the variational transfer between fractures and matrix for non-conforming methods needs to be implemented both accurately and with parallel processing capabilities.
To this end, \citet{krause_2016} developed a general, accurate and parallel variational transfer approach, which requires no prior knowledge about the relationship between the two meshes.
More specifically, an $L^2$-projection variational transfer operator has been shown \cite{hesch} to provide better approximations than interpolations. 
Previous applications of this $L^2$-projection include fluid--structure interaction (FSI) problems and mechanical contact of rough fractures \cite{nestola_2017,vonPlanta_2018,vonPlanta_2018b,vonPlanta_2018c}.
Typically, these applications are solved in equi-dimensional domains.
To address the representation of fractures by lower-dimensional manifolds (i.e.\ surface elements), the $L^2$-projection algorithm has been extended to surface--volume interactions.
This enables transfer of information between surface elements and volume elements for fractures and matrices, respectively.
Building on the work of  \citet{krause_2016} and \citet{koeppel_2018a}, the aim of this work is to demonstrate their methods' applicability to steady-state, single-phase fluid flow in 3D fractured porous media.\\
\\
%%% FINDINGS
This paper presents an application of the Lagrange multiplier method in combination with the $L^2$-projection variational transfer operator in 3D. 
Section~\ref{sec:methods} provides a brief overview of the method, by discussing the mathematical formulation, the discretization, the surface--volume interaction, and the implementation. 
Section~\ref{sec:numericalResults} first compares 2D and 3D results to state-of-the-art benchmark results \cite{flemisch_2018}. 
Next, the method is validated for 3D fracture networks by comparing it to results from conforming mesh simulations, which are used as a reference. 
For all cases described above, different mesh sizes and mesh convergences are discussed.
Finally, we apply the method to realistic fracture networks with hundreds of fractures, demonstrating the capability of the Lagrange multiplier method to model single-phase flow through 
realistic 3D fracture networks. 
The presented findings are then summarized in Section~\ref{sec:conclusion}.
%%%
%%%
%%% METHODS
%%%
%%%
\section{Method} \label{sec:methods}
In order to accommodate fluid flow through 3D fractured porous rock volumes, the Lagrange multiplier formulation, proposed by \citet{koeppel_2018a}, is applied and solved in 3D.
The Lagrange multiplier formulation considers fractures as lower-dimensional manifolds (i.e.\ surface elements). 
In 3D, this results in surface domains for the fractures and one volume domain for the rock matrix.
Accurate transfer of fluid pressure between surface and volume domains is accomplished by using the $L^2$-projection variational transfer operator \cite{krause_2016} to discretize the Lagrange multipliers. 
The following subsections discuss the mathematical formulation, followed by a description of the discretization, 
the $L^2$-projection method for surface--volume interaction, and the implementation.

%%%
%%% SUBSEC _ MATH DESCRIPTION
%%%
\subsection{Mathematical formulation}

\begin{figure}[h]
    \centering
    \begin{tikzpicture}
		\node[anchor=south west,inner sep=0] at (0,0) {\includegraphics[width=.5\linewidth]
        {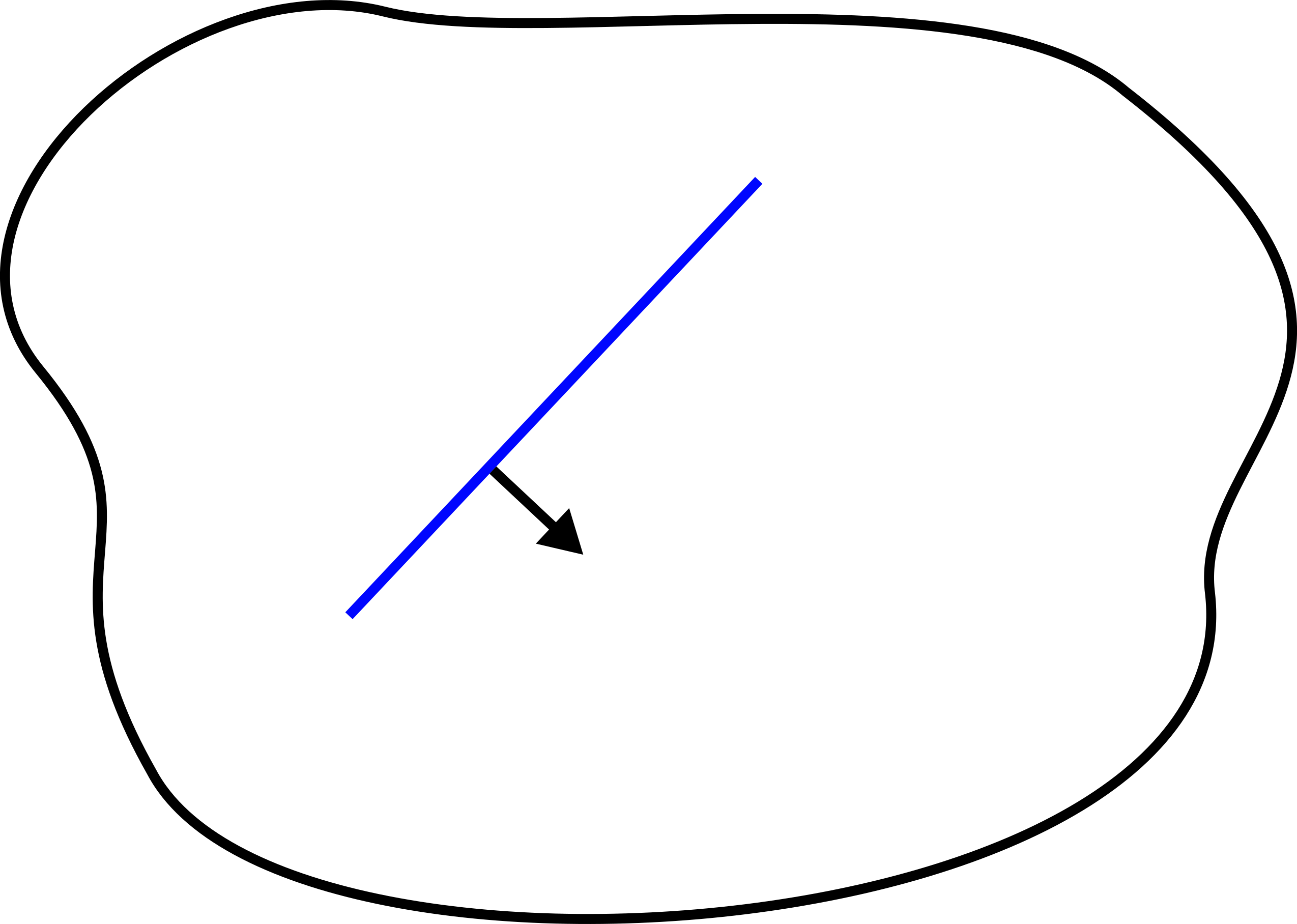}};
		\draw (5.5,3) node[font=\normalsize]{$\Omega$};
        \draw (3.5,4) node[font=\normalsize]{$\gamma$};
        \draw (3,1.7) node[font=\normalsize]{$\mathbf{n}_{\gamma}$};
	\end{tikzpicture}
    \caption{2D matrix domain $\Omega$ with an embedded 1D fracture domain $\gamma$ and normal vector $\mathbf{n}_{\gamma}$ on $\gamma$.}
    \label{fig:fractureExample}
\end{figure}
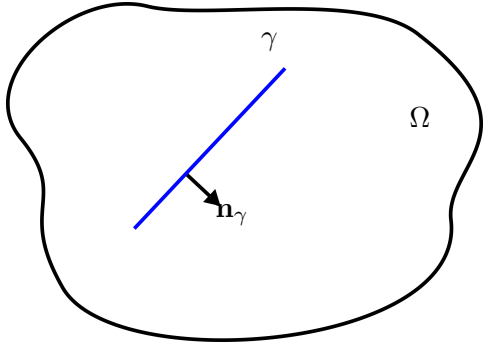

Following \citet{koeppel_2018a}, we formulate a continuous Lagrange multiplier fracture problem. 
\\
The matrix domain is designated with $\Omega$ $\subset$ $\mathbb{R}^n$, $n=2$ or $3$, and the fracture domain with 
$\gamma$ $\subset$ $\Omega$ of dimension $n-1$. 
A normal vector, $\textbf{n}_{\gamma}$, is defined with respect to the fracture surface (Fig.~\ref{fig:fractureExample}).
Steady-state fluid flow in the porous-medium matrix, $\Omega$, is governed by
\begin{equation}
    \label{eq:1}
    \begin{aligned}
    \nabla\cdot(-\mathbf{K} \nabla p)-\lambda=f & \qquad \text{in}\quad\Omega\,, \\
    p=0 & \qquad \text{on}\quad \Gamma =\partial \Omega\,,
    \end{aligned}
\end{equation}
where $\mathbf{K}$ is the permeability tensor, $p$ is the fluid pressure, and $f$ is the sink/source term.\\
Flow in the fracture, $\gamma$, is described by
\begin{equation}
    \label{eq:2}
    \begin{aligned}
    \nabla_{\gamma}\cdot(-\mathbf{K}_{\gamma} \nabla_{\gamma} p_{\gamma})+\lambda=f_{\gamma} & \qquad \text{in}\quad\gamma \,,\\
    p_{\gamma}=0 & \qquad \text{on} \quad \Gamma =\partial \gamma \,.
    \end{aligned}
\end{equation}
Above, $\partial\Omega$ and $\partial\gamma$ is the interface boundary between $\Omega$ and $\gamma$.
Fluid exchange between $\Omega$ and $\gamma$ is given by $\lambda=\lambda(x),\; x\in \gamma$.
\\
The spaces $V_{\Omega},V_{\gamma}, \mathbf{V}$, and $\Lambda$ are defined by:
\begin{equation}
    \label{eq:3}
    \begin{aligned}
    V_{\Omega}=H_0^1(\Omega), & \qquad V_{\gamma}=H_0^1(\gamma),\\
    \mathbf{V}=V_{\Omega}\times V_{\gamma}, & \qquad \Lambda=H_{0,0}^{\frac{1}{2}}(\gamma),
    \end{aligned}
\end{equation}
with the test functions $q \in V_{\Omega}$, $q_{\gamma} \in V_{\gamma}$, and $\mu \in \Lambda$.
The variational formulation is found by multiplying Eqs.~\eqref{eq:1} and~\eqref{eq:2} by the test functions, integrating over $\Omega$ and $\gamma$, and using integration by parts on both equations.
From that, the variational formulation is given as follows:\\ \medskip
Find $(p,p_{\gamma})\in \mathbf{V}$ and $\lambda \in \Lambda$, such that 
\begin{equation}
    \label{eq:4}
    \begin{aligned}
    \int_\Omega\mathbf{K}  \nabla p\cdot\nabla q+\int_{\gamma}\mathbf{K}_{\gamma}  
    \nabla_{\gamma}p_{\gamma}\cdot\mathbf{\nabla_{\gamma}}q_{\gamma}-\\
    \int_\gamma \lambda(q-q_\gamma)=\int_\Omega fq+\int_\gamma f_\gamma q_\gamma\,, & \qquad \forall(q,q_\gamma)\in\mathbf{V}
    \end{aligned}
\end{equation}
and
\begin{equation}
    \label{eq:4b}
    \begin{aligned}
    \int_\gamma (p-p_\gamma)\mu = 0\,, & \qquad \forall\mu\in\Lambda \, .
    \end{aligned}
\end{equation}
Here, Eq.~\eqref{eq:4b} indicates the coupling conditions between the domains $\Omega$ and $\gamma$.
The Lagrange multiplier represents the fluid pressure gradient $\lambda=\mathbf{K}\nabla p\cdot\mathbf{n}_\gamma$ and ensures the continuity of the fluid pressure and the exchange of the forces between the fracture domain, $\gamma$, and the matrix, $\Omega$, in the direction normal to $\gamma$.\\
\citet{koeppel_2018a} show that there exists a unique solution to the variational formulation of this problem. 
Further details on the mathematical proof can be found in their work.
%
%%%
%%% FIG - FRACTURE EXAMPLE
%%%
%
%
%%%
%%% SUBSEC - DISCRETIZATION
%%%
\subsection{Discretization}
Based on the variational formulation in Eq.~\eqref{eq:4}, the discrete counterpart is formulated in a finite element framework.
In order to solve the discrete formulation, three distinct meshes are defined to approximate the 
matrix $\Omega$, the fracture $\gamma$, and the Lagrange multiplier $\lambda$ in the fracture.\\
The meshes for the finite element discretization of Eq.~\eqref{eq:4} are defined to be $\mathcal{M}$ in $\Omega$, 
$\mathcal{M}_{\gamma}$ in $\gamma$, and $\mathcal{M}_{\lambda}$ in $\lambda$.
The respective mesh widths $h_{\mathcal{M}}$, $h_{\mathcal{M}, \gamma}$, and $h_{\mathcal{M}, \lambda}$ are defined by:
\begin{itemize}
\item[] $h_{\mathcal{M}}:= \max\limits_{1 \leq M \leq \mathcal{M}}h_M$, where $h_M$ = diam $M$, 
\item[] $h_{\mathcal{M,\gamma}}:= \max\limits_{1 \leq m \leq \mathcal{M}_{\gamma}}h_m$, where $h_m$ = diam $m$,
\item[] $h_{\mathcal{M,\lambda}}:= \max\limits_{1 \leq n \leq \mathcal{M}_{\lambda}}h_n$, where $h_n$ = diam $n$.
\end{itemize}
To enhance readability, the respective mesh widths are reduced to $h$, $h_{\gamma}$, and $h_{\lambda}$ for the 
remainder of this paper.\\
The shape of elements in each mesh might be of any kind. 
Hence, the approximation spaces $V_{h,\Omega}$, $V_{h,\gamma}$, and $\Lambda_{h}$ of continuous, 
piecewise-polynomial functions on $\Omega$, $\gamma$, and $\lambda$ are defined by:
\begin{equation}
    \label{eq:discrOG}
    \begin{aligned}
    V_{h,\Omega}=&\{q\in H_{0}^{1}(\Omega):\forall M \in \mathcal{M},\\
    & q|_M \in
    \left\{\begin{aligned}
    \mathbb{P}^2(M)     &\:\textrm{if }M\textrm{ is: a triangle, pyramid, tetrahedron, or prism}\\
    \mathbb{Q}^{2,2}(M) &\:\textrm{if }M\textrm{ is: a quadrilateral or hexahedron}\\
    \end{aligned}
    \right\},\\
    V_{h,\gamma}=&\{q_{\gamma}\in H_{0}^{1}(\gamma):\forall m \in \mathcal{M}_{\gamma},\\
    & q_{\gamma}|_m \in
    \left\{\begin{aligned}
    \mathbb{P}^1(m)     &\:\textrm{if }m\textrm{ is: a triangle or line segment}\\
    \mathbb{Q}^{1,1}(m) &\:\textrm{if }m\textrm{ is: a quadrilateral}\\
    \end{aligned}
    \right\},\\
    \Lambda_{h}=&V_{h,\gamma}.&\\
    \end{aligned}
\end{equation}
The definition of the approximation space $\Lambda_{h}$ of the Lagrange multiplier implies that $h_{\lambda}=h_{\gamma}$.
To prevent a poorly conditioned system matrix, it is necessary to satisfy $h_{\lambda} \leq \min(h, h_{\gamma})$, which -- in
combination with the above -- results in $h\leq h_{\gamma}$. 
Generally, by using $q_{\gamma}|_m \in\mathbb{P}^1(m)$ and $\lambda|_n \in\mathbb{P}^1(n)$, the Lagrange multiplier $\lambda$ 
is applied on the fracture mesh. 
This offers higher flexibility regarding the meshing.
%
%%%
%%% SUBSEC - SURFACE-VOLUME INTERACTION
%%%
\subsection{Volume--surface information transfer}
\begin{figure}[t]
    \centering \footnotesize
    \includegraphics[width=.98\linewidth]{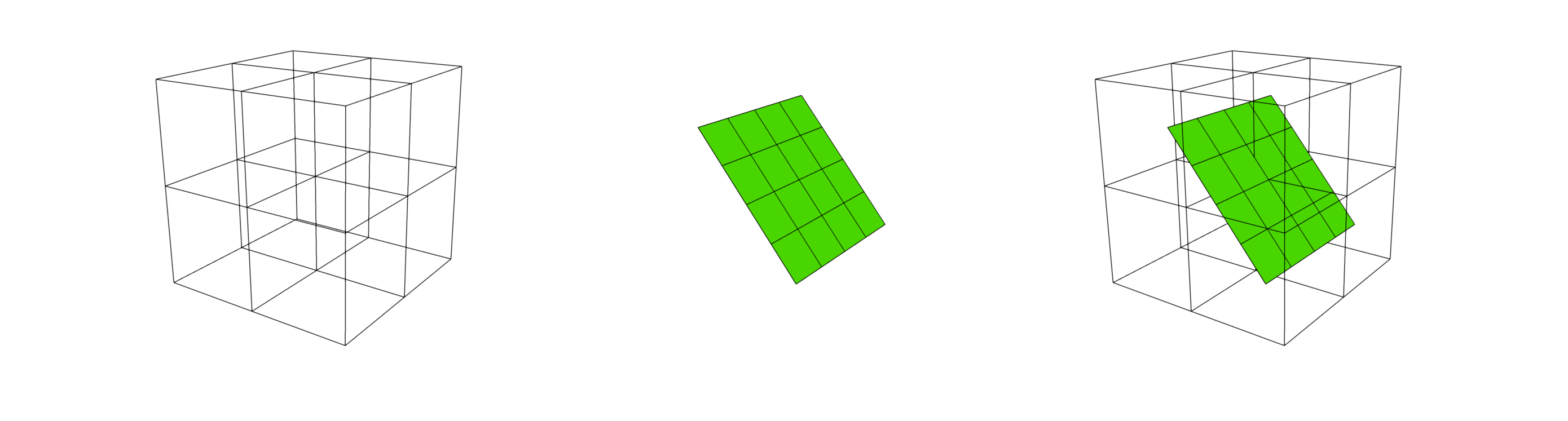}
    \caption{Examples of matrix volume mesh (left) and inclined fracture surface mesh (center). The two meshes are combined in a single model (right).}
	\label{fig:meshexamples}
\end{figure}

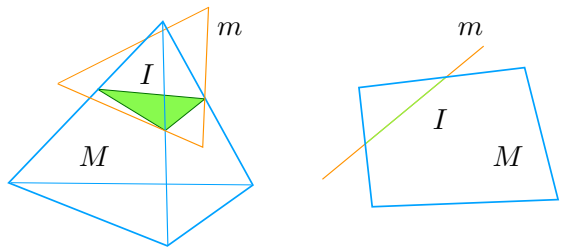
\begin{figure}[b]
    \centering
    \begin{tikzpicture}
		\node[anchor=south west,inner sep=0] at (0,0) {\includegraphics[width=.6\linewidth]
        {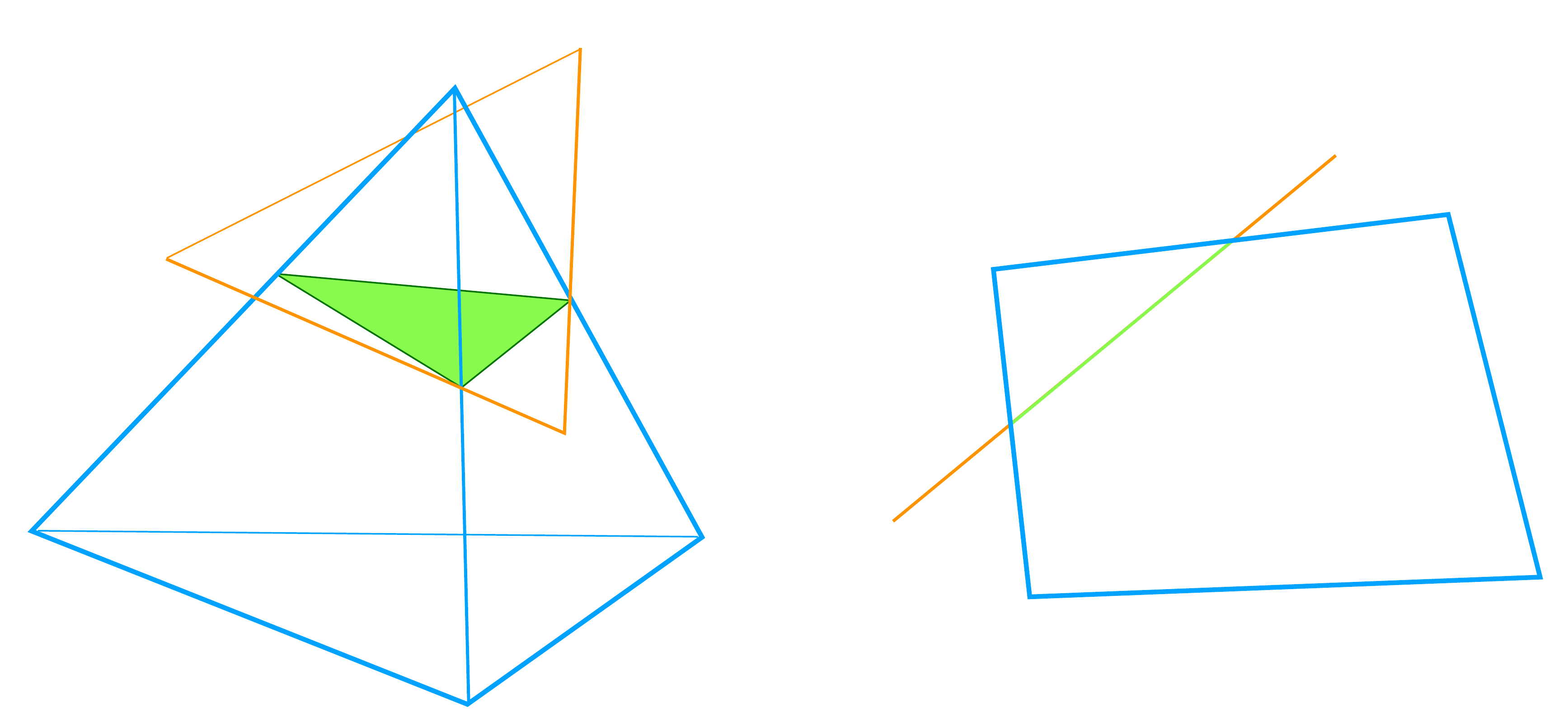}};
        % 3D
		\draw (1.3,1.3) node[font=\normalsize]{$M$};
        \draw (2.0,2.4) node[font=\normalsize]{$I$};
        \draw (3.1,3.0) node[font=\normalsize]{$m$};
        % 2D
		\draw (6.8,1.3) node[font=\normalsize]{$M$};
		\draw (5.9,1.8) node[font=\normalsize]{$I$};
        \draw (6.3,3.0) node[font=\normalsize]{$m$};
	\end{tikzpicture}
    \caption{Example of intersections $I$ between elements of meshed domains $M$ and $m$. Left: 3D. Right: 2D.}
    \label{fig:intersectionExample}
\end{figure}

The meshes associated with the spaces $V_{h,\Omega}$ (matrix) and $V_{h,\gamma}$ (fracture network) are generally non-matching (i.e.\ Fig.~\ref{fig:meshexamples}), which means that the surfaces of the volume elements of the matrix do not necessarily coincide with those of the surface elements of the fracture network. 
%%%
%%% FIG - EXAMPLE INTERSECTING MESH DOMAINS
%%%
This leads to mutually non-conforming discretizations which require the use of information transfer techniques such as interpolation or $L^2$-projections for handling the coupling terms in Eq.~\eqref{eq:4}. 
Here the $L^2$-projection approach is adopted, as it has been shown to have better approximation properties than interpolation~\cite{hesch}.\\
Ultimately, intersections between the matrix mesh and the fracture mesh have to be found in order to perform quadrature on the coupling term in Eq.~\eqref{eq:4b} up to numerical precision.
Then, integration is done on these intersections.
For any pair of a volume element $M \in \mathcal{M}$ and a shell element $m \in \mathcal{M}_\gamma$, the following procedure is completed:
\begin{itemize}
	\item Computation of the intersection $I = M \cap m$ by using a variant of the Sutherland--Hodgman clipping algorithm~\citep{sutherland1974} as shown in Fig.~\ref{fig:intersectionExample}. 
    Here, $M$ is interpreted as a set of half-spaces which are sequentially used to clip $m$.  
	\item If $I \neq \emptyset$, $I$ is meshed into the simplicial complex $\mathcal{T}_I = \{ S \}$, where $S$ is a simplex.
    \item A suitable quadrature rule is mapped to each simplex $S$ which is then mapped to the reference configurations of $M$ and $m$.
\end{itemize}
%%%
%%% FIG - EXAMPLE OF INTERSECTION
%%%
The task of detecting pair-wise element intersections is accelerated by employing octree data structures. 
Additional acceleration can be gained by applying techniques such as spatial-hashing~\cite{realtimecol}.
The applied algorithms are fully automated and no prior knowledge about the relation of the meshes is required. 
Further details regarding the information transfer procedure can be found in~\citet{krause_2016}.
%
%%%
%%% SUBSEC - IMPLEMENTATION
%%%
%
\subsection{Implementation}
The routines described in this paper are implemented within the open-source software library \emph{Utopia}~\cite{utopiagit}. \emph{Utopia} uses \emph{libMesh}~\cite{libmesh} for the finite element discretization, \emph{MOONoLith}~\cite{moonolith} for the intersection detection, and \emph{PETSc}~\cite{petsc} with \emph{MUMPS}~\cite{mumps} for the linear algebra calculations. 
In the numerical experiments illustrated in Section~\ref{sec:numericalResults}, the size of the algebraic linear system of equations~\eqref{eq:4} reaches at most one million degrees of freedom, which is solved with the \emph{MUMPS} direct solver.
%
%%%
%%%
%%% SEC - NUMERICAL RESULTS
%%%
%%%
\section{Numerical results \& Discussion} 
\label{sec:numericalResults}
%%% OVERVIEW
This section discusses various numerical experiments performed to investigate the Lagrange multiplier method in 2D and 3D.
First, the implementation is verified in 2D by comparing the results to a benchmark case presented by
\citet{flemisch_2018}.
In a second experiment, the 2D case is extruded in the third dimension.
This enables testing the accuracy of the Lagrange multiplier method, combined with the $L^2$-projection variational transfer operator (LM--L$^2$) in 3D by comparing it to the 2D benchmark results.
Next, a heterogeneous 3D fracture network is built and the LM--L$^2$ method is validated by comparing it with results from a conforming mesh model.
Finally, the LM--L$^2$ method is applied to a realistic fracture network with 150 randomly distributed fractures. 
These final numerical experiments are conducted to investigate different geometrical complexities as well as mesh convergence. \\
%%% MESH
Throughout the numerical experiments, various element shapes are used, following Eqs.~\eqref{eq:discrOG}.
The implementation of the LM--L$^2$ method causes no restrictions regarding the element shape, so that the matrix domain, $\Omega$, and the fracture domain, $\gamma$, can therefore be of arbitrary shape. 
However, for simplicity, the matrix mesh, in all presented numerical experiments, is composed of hexahedral (for 3D) and quadrilateral (for 2D) elements. 
The respective element spaces for all experiments is second order in $\Omega$ and first order in $\gamma$ and $\lambda$. Nevertheless, the implementation allows consideration of zero-, first-, and second-order elements in all domains. 
To test convergence and accuracy, various mesh sizes for $\Omega$ and $\gamma$/$\lambda$ are employed throughout this study. \\
%%% PARAMETERS
To facilitate comparison with existing studies \citep{flemisch_2018,koeppel_2018a}, all physical parameters are normalized throughout this study. 
All experiments are conducted on a unit domain with an edge length of $1$. 
Further, the fracture permeability is $\mathbf{K}_{\gamma}=10^4\mathbf{I}_{\gamma}$ and the fracture aperture is $a=10^{-4}$, which is incorporated in the applied fracture permeability and the Neumann boundary condition.
The permeability in the matrix domain is set to $\mathbf{K}=\mathbf{I}$.
These choices ensure that, for the cases studied here, the matrix and fracture network both contribute to a similar extent to the overall fluid flow, i.e.\ the ratio between a) the average aspect ratio of the fractures and b) the permeability ratio between the matrix and the fractures approximately equal to one \cite{ebigbo_2016}. 
%
%
%%%
%%% SUBSEC - BENCHMARKS 2D
%%%
\subsection{Benchmark -- 2D} \label{sec:benchmark2D}
\begin{figure}[h]
    \centering
    \includegraphics[width=.8\linewidth]{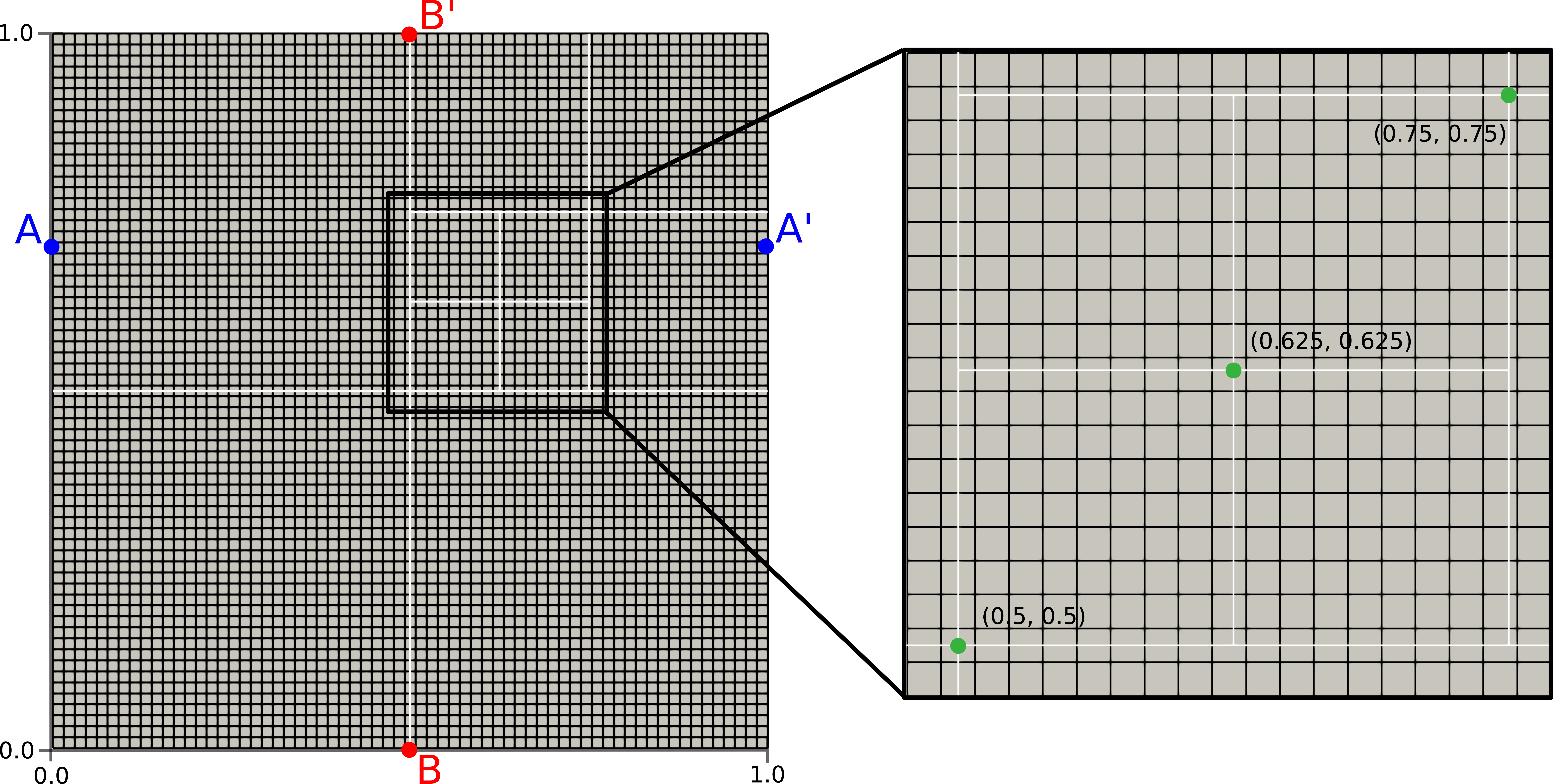}
    \caption{
    Numerical mesh used for the 2D benchmark case \cite{flemisch_2018}.
    The fracture network is shown in white. The enlarged region on the right shows the fracture intersections. 
    The displayed mesh has a discretization of $h=1/65$.
    Points A, A$^{\prime}$, B and B$^{\prime}$ mark the observation lines AA$^{\prime}$ and
    BB$^{\prime}$ along which fluid pressure profiles are plotted.
    }
    \label{fig:benchmark2Da}
\end{figure}

\begin{figure}[h]
    \centering
    \includegraphics[width=0.6\linewidth]{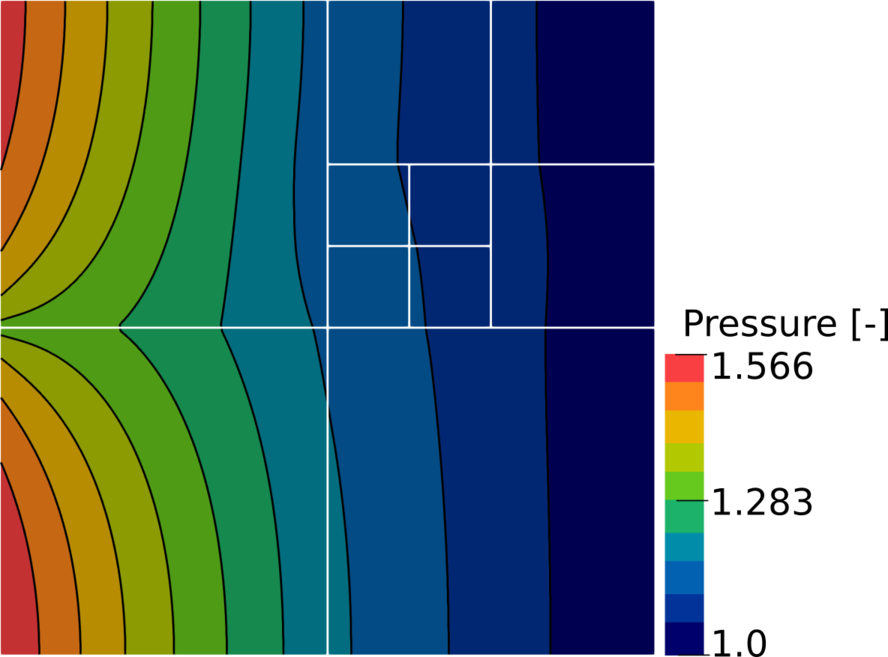}
    \caption{2D benchmark case, consisting of a pressure gradient from left to right and six embedded, 
    lower-dimensional fractures with higher permeability \cite{flemisch_2018}.
    The discretization is $h=1/129$ and $h_{\gamma}=h_{\lambda}=1/128$.}
    \label{fig:benchmark2Db}
\end{figure}

\begin{figure}[h]
    \centering
    \includegraphics[width=0.99\linewidth]{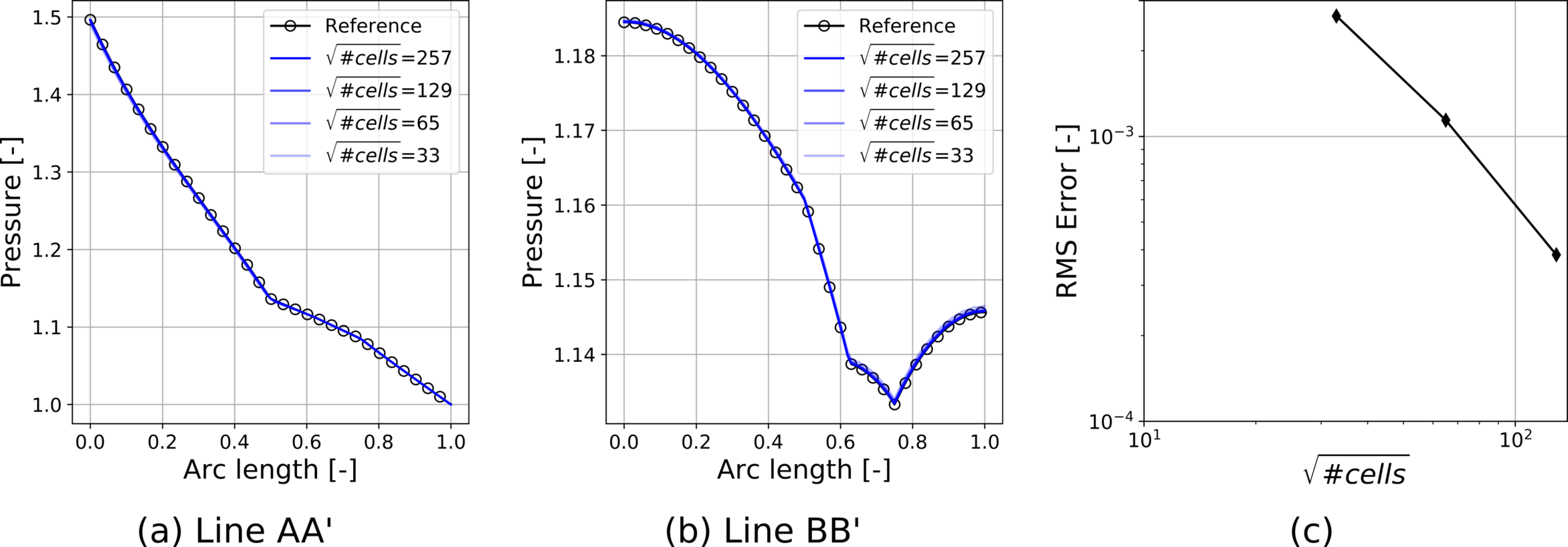}
    \caption{
    Numerical convergence with mesh refinement for the 2D benchmark case \cite{flemisch_2018}. 
    Shown are: (a) the fluid pressure profile along the line AA$^\prime$ (see Fig.~\ref{fig:benchmark2Da}); (b) the pressure profile along the line BB$^\prime$ (see Fig.~\ref{fig:benchmark2Da}); and c) the RMS$_{\mathcal{M}}$ error for $h=1/33$, $h=1/65$, and $h=1/129$, relative to $h=1/257$ and $h_{\gamma}=h_{\lambda}=1/256$ (see Eq.~\ref{eq:RMS}).
    }
    \label{fig:benchmark2Dc}
\end{figure}

%%% OVERVIEW
First, the implementation of the LM--L$^2$ method is tested by comparing the obtained results to 2D benchmark results presented by \citet{flemisch_2018}. 
Benchmark~$1$, which was first introduced by \citet{geiger_2013}, is chosen as a representative example for this study.
Emphasis is placed on the validation of the presented method with reference results from \citet{geiger_2013,flemisch_2018}, while results of alternative approaches can be found in the listed references.
While \citet{koeppel_2018a} used $\mathbb{P}^0$ elements for their benchmark comparison, this study employs $\mathbb{P}^1$ elements for the Lagrange multiplier mesh and $h_{\gamma}=h_{\lambda}$. 
The fracture elements consist of line segments and the matrix of quadrilateral elements.
Further, a mesh-size convergence study is conducted for the fracture and the matrix mesh. \\
%%% BENCHMARK SETUP
The fracture network at hand contains six fractures in perpendicular orientation as shown in Fig.~\ref{fig:benchmark2Da}. 
%%%
%%% FIG - BENCHMARK CASE 2D MESH 
%%%
A non-homogeneous Neumann boundary condition on the left domain boundary serves as a fluid source and a non-homogeneous Dirichlet boundary condition on the right domain boundary serves as a fluid sink, which yields fluid flow across the domain from the left to the right boundary.
Homogeneous Neumann boundary conditions at the top and bottom boundaries enforce no-flow conditions across those domain boundaries. \\
%%% BENCHMARK RESULTS
The resulting pressure distribution for $h=1/129$, and $h_{\gamma}=h_{\lambda}=1/128$ is depicted in Fig.~\ref{fig:benchmark2Db}, and shows good agreement with the reference benchmarks \citep{geiger_2013,flemisch_2018}. 
%%%
%%% FIG - BENCHMARK CASE 2D CONTOUR
%%%
As expected, the central, horizontal fracture serves as a channel of high permeability, facilitating faster fluid flow than the surrounding porous-medium matrix. 
Matrix regions further removed from the high-permeability fractures (e.g., top and bottom left corner) therefore result in the steepest fluid pressure gradients. \\
%%% CONVERGENCE STUDY
Mesh-size dependency is subsequently investigated in a convergence study. 
Mesh widths are chosen such that non-conforming meshes for the fractures and the matrix are ensured.
The initial realization has the largest mesh width of $h=1/33$ and $h_{\gamma}=h_{\lambda}=1/32$.
With each refinement, the mesh resolution is then increased by a factor of two, resulting in 
$h=1/65$ and $h_{\gamma}=h_{\lambda}=1/64$, $h=1/129$ and $h_{\gamma}=h_{\lambda}=1/128$, and $h=1/257$ and $h_{\gamma}=h_{\lambda}=1/256$. \\
%%% CONVERGENCE STUDY RESULTS
Figs.~\ref{fig:benchmark2Dc}a and \ref{fig:benchmark2Dc}b show the results for all realizations and the reference results along the lines AA$^{\prime}$ and BB$^{\prime}$, yielding good convergence and accuracy for all realizations. 
%%%
%%% FIG - BENCHMARK CASE 2D COMBINED FIGURE
%%%
However, for low resolutions of $h_{(\gamma/ \lambda)}$, small deviations can be observed along BB$^{\prime}$ between $Arc\ length=0.6$ and $Arc\ length=1.0$. 
This can be attributed to a lack in resolution of the fracture intersections by the non-conforming mesh configuration. 
This error increases with decreasing matrix mesh resolution.
\citet{koeppel_2018a} suggest that results might be more accurate for $h\geq h_{\gamma,\lambda}$.
However, the considered second-order function space in $\Omega$ reduces this effect. \\
%%% RMS
Fig.~\ref{fig:benchmark2Dc}c shows the RMS$_{\mathcal{M}}$ in the matrix for $h=1/33$, $h=1/65$, and $h=1/129$, 
relative to $h=1/257$, which is decreasing approximately linearly with increasing mesh resolution. 
The error between different results is calculated by the root mean square (RMS) over all elements.
To facilitate comparison of non-conforming meshes, results from all resolutions are interpolated on a mesh with the finest resolution.
The squared error on a single element in the mesh $\mathcal{M}_{(\gamma)}$ and the RMS$_{\mathcal{M},(\gamma)}$ error are calculated following:
\begin{equation}
    \label{eq:RMS}
    \begin{aligned}
      \text{err}_{n}^2=&\left(p_{n,\text{ref}} - p_{n}\right)^2\,, \\
      \qquad \text{RMS}_{\mathcal{M}, (\gamma)}=&\sqrt{\frac{1}{N}\sum\limits_{n=1}^{N}\text{err}_{n}^2}\,,\\
    \end{aligned}
\end{equation}
where $n$ is a node in $\mathcal{M}$ or $\mathcal{M}_{\gamma}$, and $N= \Sigma n$. 

%%%
%%% SUBSEC - BENCHMARKS 3D
%%%
\subsection{Benchmark -- 3D}
\label{sec:benchmark3D}
\begin{figure}[h]
    \centering
    \includegraphics[width=0.99\linewidth]{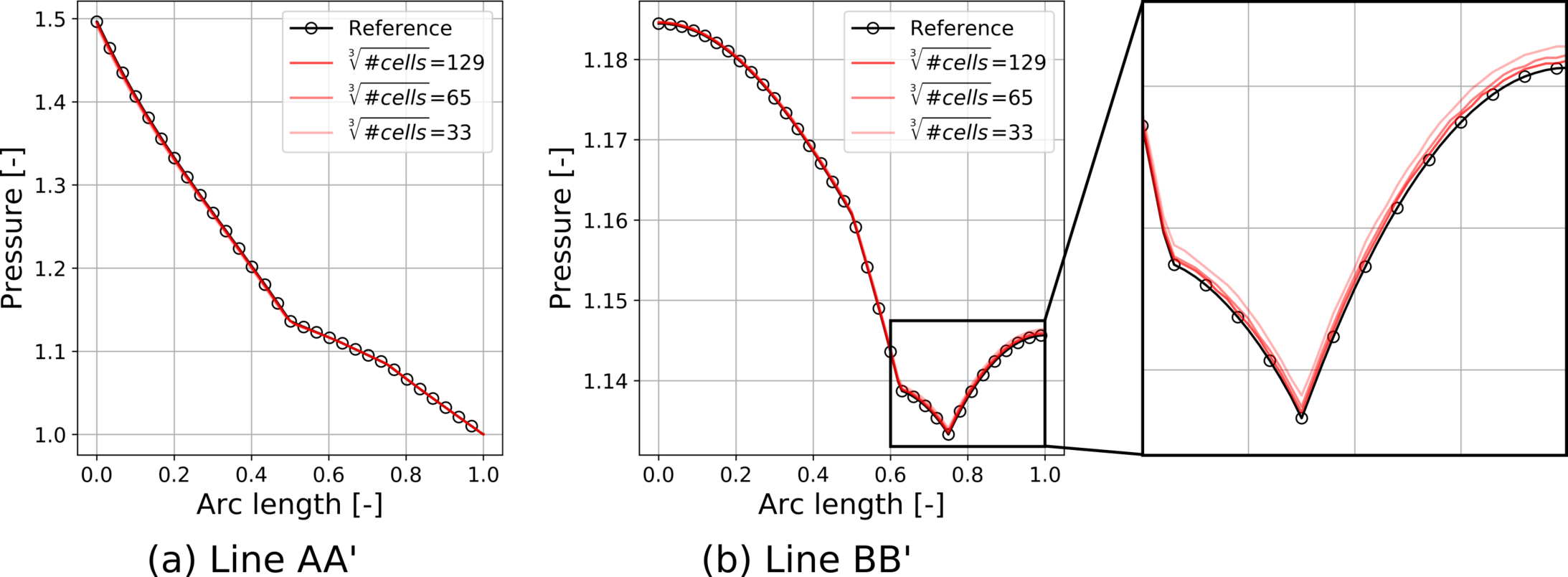}
    \caption{
    Pressure profiles along the lines of: (a) AA$^\prime$; and (b) BB$^\prime$ as shown in Fig.~\ref{fig:benchmark2Da}.
    The two reference lines are located in the center of the extrusion in the third dimension.
    Numerical convergence is shown by reference results and three mesh refinements.
    }
    \label{fig:benchmark3D}
\end{figure}

%%% SETUP
The accuracy of the LM--L$^2$ method in 3D is tested by extruding the setup introduced in Section~\ref{sec:benchmark2D} in the third dimension, which results in a 3D porous-medium matrix domain and surface domains for the fractures (Fig.~\ref{fig:benchmark2Da}). 
This extrusion enables comparison of 3D results and the 2D benchmarks from \citet{flemisch_2018}.
The extrusion causes the observation lines AA$^{\prime}$ and BB$^{\prime}$ (see Fig.~\ref{fig:benchmark2Da}) to be located anywhere in the direction of the extrusion.
For the present case, their position is chosen at half of the extrusion length.
Non-homogeneous Neumann boundary conditions are applied at the fracture and the matrix on the left boundary.
A non-homogeneous Dirichlet boundary condition is applied on the right boundary.
Homogeneous Neumann boundary conditions are imposed on the remaining boundaries. \\
%%% RESULTS INTRO
Various realizations with different mesh sizes are performed to study convergence.
Fig.~\ref{fig:benchmark3D} shows the pressure profile along the lines (a) AA$^{\prime}$ and (b) BB$^{\prime}$ for the reference results in 2D and the 3D results for all realizations.
The initial mesh width of $h=1/33$ and $h_{\gamma}=h_{\lambda}=1/32$ is consecutively refined by a factor of two, 
which results in $h=1/65$ and $h_{\gamma}=h_{\lambda}=1/64$, and $h=1/129$ and $h_{\gamma}=h_{\lambda}=1/128$.
The element shapes in this experiment are hexahedrons in $\Omega$ and quadrilaterals in $\gamma/ \lambda$.
A more detailed convergence study in 3D is presented for the more complex case in Section~\ref{sec:heteroDFN_7_fracs}. 
Here, we restrict ourselves to a comparison of the pressure results along the observation lines AA$^{\prime}$
and BB$^{\prime}$ with the reference results (Fig.~\ref{fig:benchmark3D}). \\
%%% RESULTS MAIN
While the pressure profile is captured for the mesh widths $h=1/65$ and $h_{\gamma}=h_{\lambda}=1/64$, and $h=1/33$ and $h_{\gamma}=h_{\lambda}=1/32$, they display a slight deviation between $Arc\ length=0.6$ and $Arc\ length=1.0$. 
As accuracy improves with decreasing mesh size, the realization with $h=1/129$ and $h_{\gamma}=h_{\lambda}=1/128$ only shows minor deviations from the reference results. 
The more pronounced deviations for these coarser mesh realizations can be explained with a lack of resolution around the fracture intersections in the area of $Arc\ length=0.6$ to $Arc\ length=1.0$.  
This suggests that the matrix mesh width around the fractures and in particular at the fracture intersections 
should be smaller. 
However, this effect in 3D is similar to the one in 2D systems (see Section~\ref{sec:benchmark2D}).\\
%%%
%%% FIG - BENCHMARK CASE 3D PRESSURE X - Y
%%%

%%%
%%% SUBSEC - HETEROGENESOUS DFN 7 
%%%
\subsection{Heterogeneous Fracture Network -- 3D}
\label{sec:heteroDFN_7_fracs}
\begin{figure}[h]
    \centering
    \includegraphics[width=0.6\linewidth]{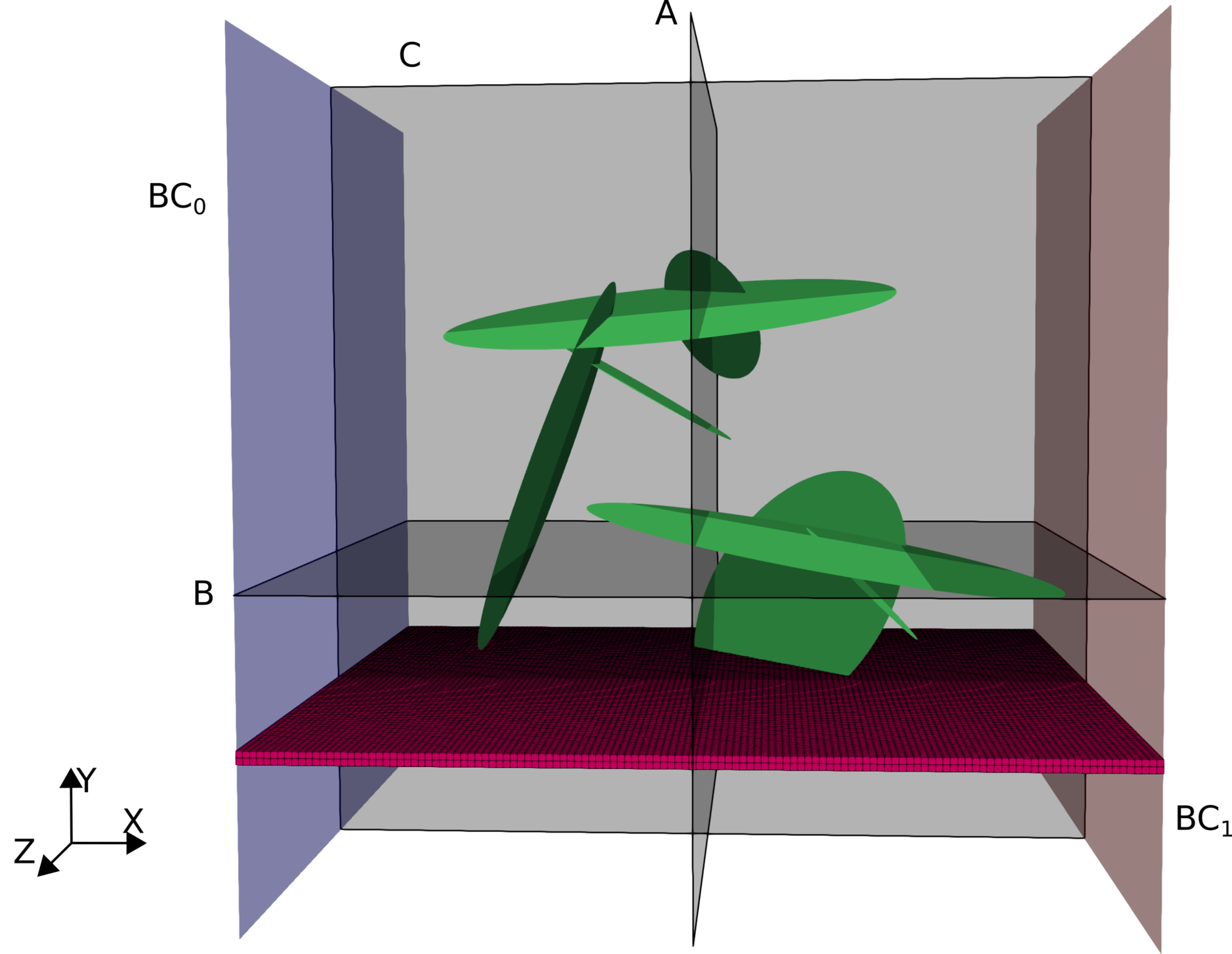}
    \caption{
    Model setup for the heterogeneous 3D case with seven embedded fractures, shown in green. 
    A slice through the porous-medium matrix mesh, with $h=1/129$, is depicted in red. 
    Dirichlet boundary conditions, BC$_0$ and BC$_1$, are located at the left and right sides of the matrix mesh.
    Three observation surfaces are located at (A) $x = 0.5$, (B) $y = 0.38$, and (C) $z = 0.5$.
    }
    \label{fig:heteroDFN_7_fracs_setup}
\end{figure}

\begin{figure}[h]
    \centering
    \includegraphics[width=0.4\linewidth]{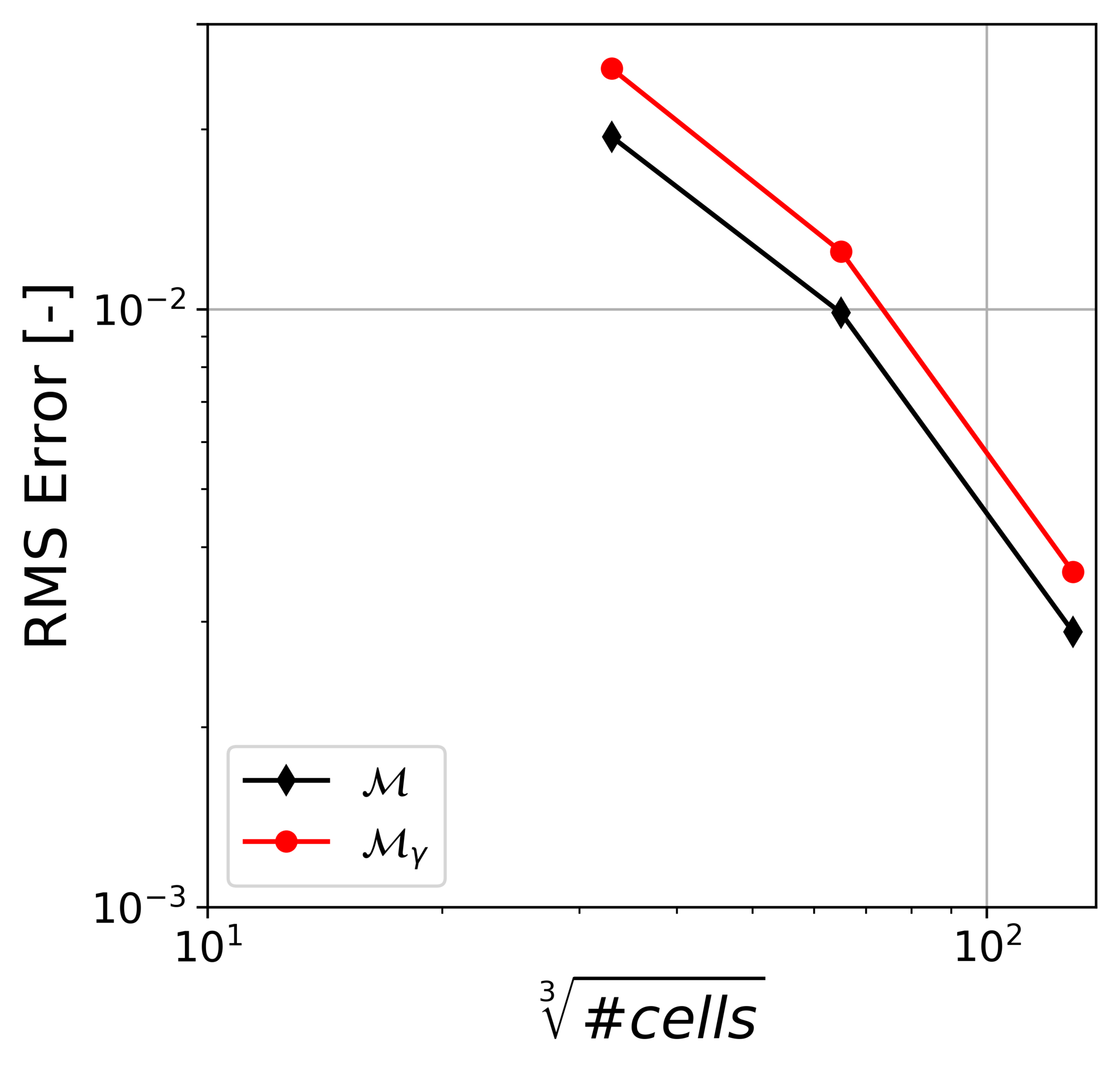}
    \caption{
    RMS error between the reference results and the results of the LM--L$^2$ method.
    The errors are calculated separately for the matrix mesh, $\mathcal{M}$, and the fracture mesh, $\mathcal{M}_{\gamma}$.
    Shown is the error for different $\sqrt[3]{\#\text{cells}}$ of the matrix mesh $\mathcal{M}$.
    }
    \label{fig:heteroDFN_7_fracs_error_matrix_fracture}
\end{figure}

\begin{figure}[h]
    \centering
    \includegraphics[width=0.99\linewidth]{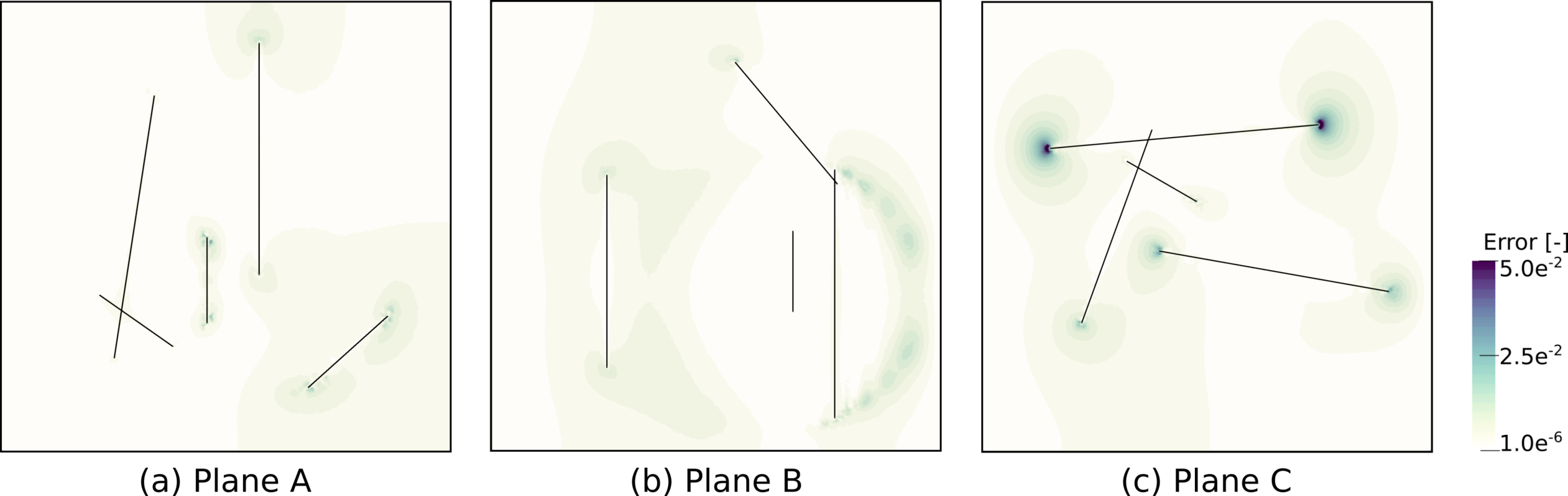}
    \caption{Error ($\sqrt{\text{err}_{\mathcal{M}}^2}$ in Eq.~\ref{eq:RMS}) between the reference results and the 
    LM--L$^2$ results for the realization 
    with $h=1/129$, $h_{\gamma}=1/129$, $h_{\lambda}=h_{\gamma}$.
    The planes A, B, and C are defined in Fig.~\ref{fig:heteroDFN_7_fracs_setup}.
    The fractures, intersected by the respective plane, are illustrated in black.}
    \label{fig:heteroDFN_7_fracs_error}
\end{figure}

%%% DFN GEOMETRIES - INTRO
Here we generate an artificial heterogeneous fracture network to test the LM--L$^2$ method with more complex geometries.
This enables the investigation of accuracy and convergence for fractures with varying orientation, size, and location.
Fractures are often assumed to be disc-shaped \cite{ebigbo_2016}. Hence, our numerical experiment also approximates fractures as circular surfaces. 
Fracture tips are generally difficult to represent by non-conforming mesh methods, as they characterize the boundary of a fracture domain.
On the tips, the Lagrange multiplier can act in multiple spatial directions, while its primary direction in the center of the fracture is normal to the fracture plane (i.e.\ flow normal to the fracture plane). 
This becomes particularly important in 3D, if the fracture edge is not a line (e.g.\ corners or circular fractures), as this results in a discontinuity of the direction of the Lagrange multiplier. 
Throughout this experiment, only the matrix-mesh width, $h$, is varied and the results are compared to 
results from a conforming mesh simulation.
This enables investigation of the influence of the matrix mesh width on the accuracy of the results. \\
%%% SETUP
The chosen model setup for seven fractures is shown in Fig.~\ref{fig:heteroDFN_7_fracs_setup}.
%%%
%%% FIG - HETEROGENEOUS DFN 7 SETUP
%%%
All fractures are located within the matrix domain and are characterized by a random diameter, orientation and location.
At the boundaries BC$_0$ and BC$_1$, the applied Dirichlet condition is set to $1$ and $0$, respectively.
The homogeneous Neumann conditions imposed on the remaining boundaries ensures no-flow. \\
%%% CONVERGENCE SETUP
Convergence is studied with three realizations with a matrix-mesh width of $h=1/33$, $h=1/65$, and $h=1/129$, respectively.
The fracture-mesh width for all realizations is constant at $h_{\gamma}=h_{\lambda}=1/129$.
The results of these realizations are compared to reference results which are obtained with a conforming mesh using finite element methods in MOOSE \cite{gaston_2009} with a mesh width of $h=h_{\gamma}=1/129$.
For the conforming mesh, the fracture elements are triangles and the matrix elements tetrahedrons.
For the LM--L$^2$ method, the element shapes in $\gamma$ are triangles and hexahedrons in $\Omega$. \\
%%% CONVERGENCE RESULTS
The observed RMS errors in the matrix and fracture show linear convergence rates with an improved rate for $h=1/129$ (see Fig.~\ref{fig:heteroDFN_7_fracs_error_matrix_fracture}).
%%%
%%% FIGURE - HETEROGENEOUS DFN 7 ERRORS IN MATRIX AND FRACTURES
%%%
Due to the similar mesh width of this realization to the reference model, this error solely compares the effect of non-conforming meshes.
Furthermore, the observed convergence rates are similar for matrix and fractures. 
Hence, the changing mesh width in the matrix mesh affects the accuracy in the fractures and the matrix similarly.
This further suggests that sufficiently high matrix resolution around the fractures is crucial.
Generally, the RMS error in the fractures is higher because the entire domain is affected by the non-matching meshes. 
In contrast, large matrix areas are further away from the fractures and therefore not affected by the non-matching meshes. \\
%%% ERROR IN CROSS SECTIONS
Fig.~\ref{fig:heteroDFN_7_fracs_error} shows the local error at cross sections A, B, and C, which are defined in 
Fig.~\ref{fig:heteroDFN_7_fracs_setup}. 
%%%
%%% FIG - HETEROGENEOUS DFN 7 FRACS
%%%
Overall, the error in planes A and B is small, although an increased error is present around the fractures and in particular the fracture tips. 
The biggest errors can be found on plane C, specifically at the fracture tips of the upper most large fracture. 
This fracture is cut through its center and has the largest diameter, thereby connecting high-pressure areas on the left with low-pressure areas on the right. 
These relatively large errors can also be observed for other fractures oriented parallel to the gradient, as they form a shortcut for pressure in the system. 
This leads to the largest pressure gradients at the fracture tips which are located close to the inflow and outflow of the domain, or which happen to be the farthest apart in the direction of the gradient. 
This is also where the biggest errors in the fracture domain are found.
% MESHING
The difficulties encountered during the generation of the conforming mesh for the reference solution serve as further motivation for the development of non-conforming methods. 
Particularly, very small elements are necessary to represent the matrix between the fracture intersections with low intersection angle. 
Specialized meshing tools are required to efficiently mesh high-quality, conforming mixed-dimensional meshes \cite{cacace_2015,holm_2006,blessent_2009}.

%%%
%%% SUBSEC - RANDOM DFN
%%%
\subsection{Random Fracture Network -- 3D}
\label{sec:randomDFN_150_fracs}
\begin{figure}[h]
    \centering
    \includegraphics[width=0.99\linewidth]{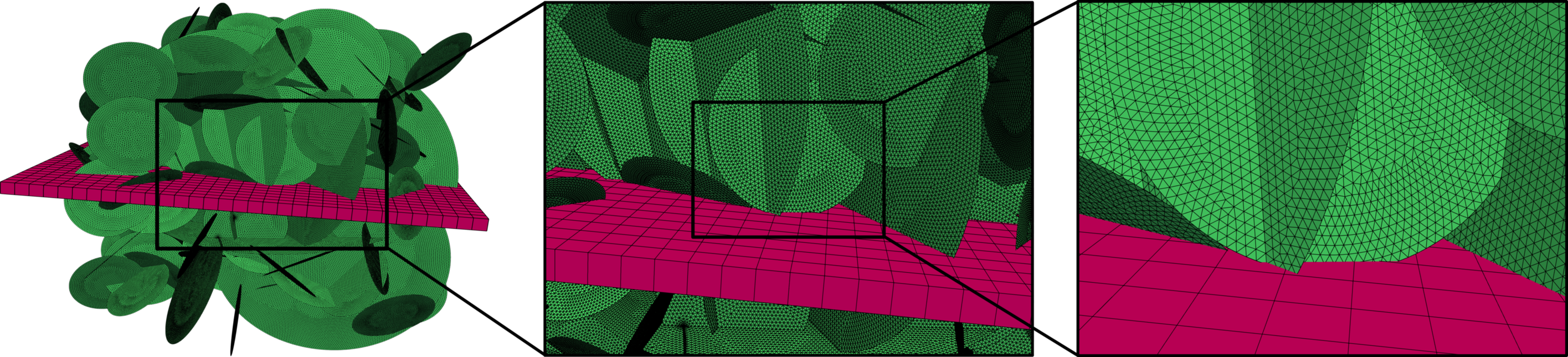}
    \caption{Mesh for the random fracture network. Shown in red is one layer of the matrix mesh. The 150 
    fractures are depicted in green. 
    The displayed mesh discretization is $h=1/33$ and $h_{\gamma}=h_{\lambda}=1/200$.
    }
    \label{fig:randomDFN_150_fracs_mesh}
\end{figure}

\begin{figure}[h]
    \centering
    \includegraphics[width=0.6\linewidth]{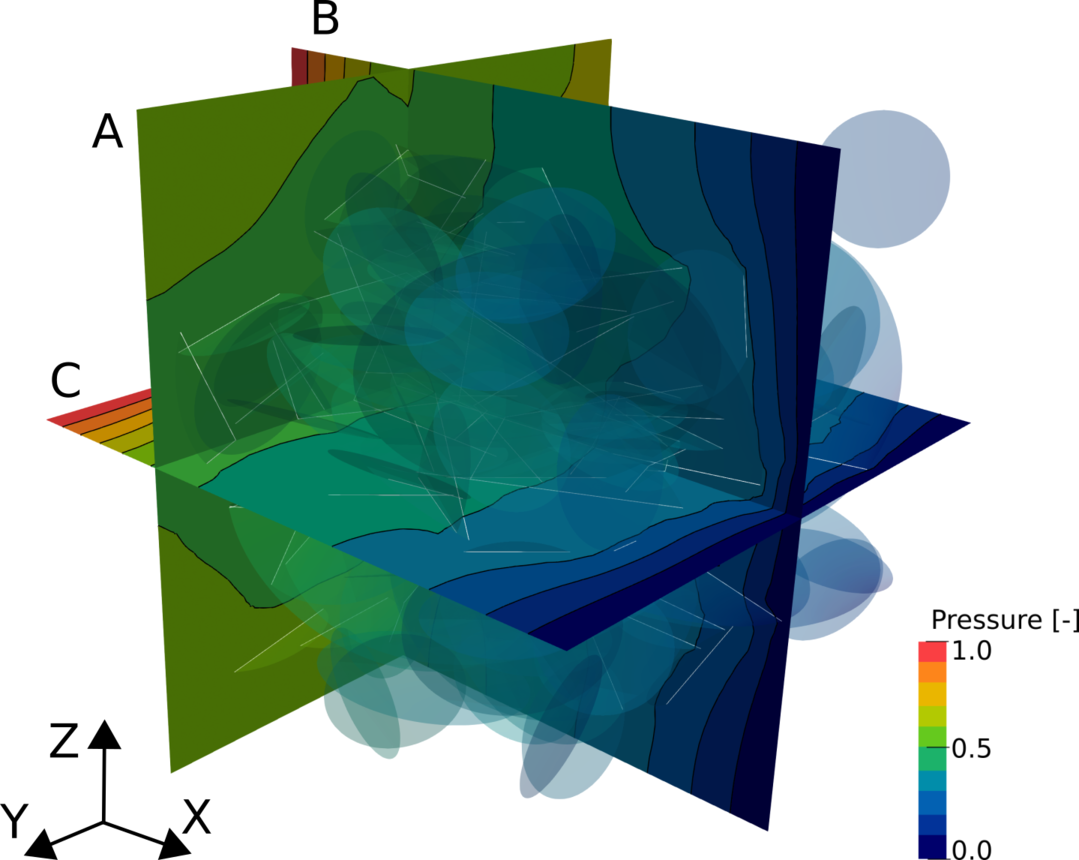}
    \caption{Dimensionless fluid pressure field in the fracture network and the porous-medium matrix. Cross sections through the matrix are given at (A) $x = 0.26$, (B) $y = 0.5$, and 
    (C) $z = 0.5$.}
    \label{fig:randomDFN_150_fracs_results3D}
\end{figure}

\begin{figure}[h]
    \centering
    \includegraphics[width=0.99\linewidth]{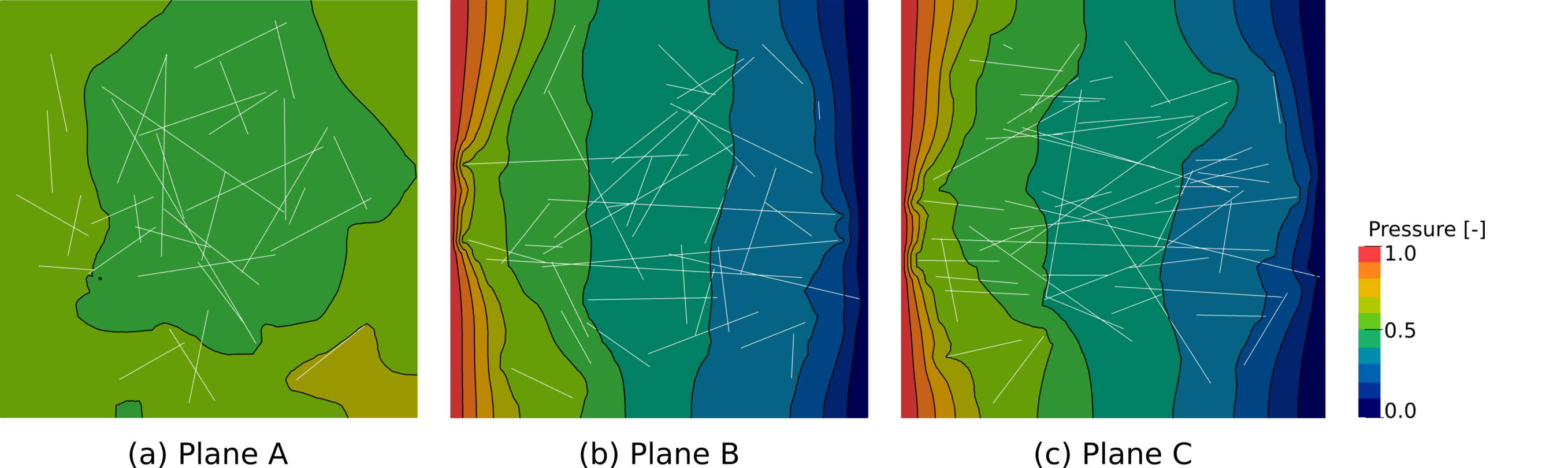}
    \caption{Pressure contours in the porous-medium matrix within Planes A, B, and C, shown in Fig.~\ref{fig:randomDFN_150_fracs_results3D}. 
    Fractures intersecting a plane are shown in white.}
    \label{fig:randomDFN_150_fracs_results2D}
\end{figure}

%%% INTRO
Fracture networks commonly contain hundreds or even thousands of fractures within a rock volume of \SI{100}{m} side length. 
They are geometrically characterized by several distributions regarding the fracture size, orientation, density or location.
It is commonly assumed that the fracture size is distributed following a truncated power law \cite{bonnet_2001}.
Fracture orientation, density, and location are site-specific and usually derived from geological information. \\
%%% LML
These highly heterogeneous geometries are computationally demanding, when generating meshes for numerical simulations. 
Using the LM--L$^2$ method, the geometry only needs to be represented by the fracture mesh, as the porous-medium matrix is represented by a
regular grid.
This numerical experiment shows and investigates the application of the LM--L$^2$ method on a
realistic fracture network with 150 fractures.\\
%%% SETUP
In this experiment, the fracture radius distribution follows a power law, with truncations at 0.1 and 0.4.
The 150 fractures are circular, randomly oriented, and distributed in a cubical model domain with a side length of 1. 
For simplicity, it is further ensured that no fracture intersects the model domain boundary.
Fig.~\ref{fig:randomDFN_150_fracs_mesh} shows the fractures embedded in the porous-medium matrix, of which only a single layer is displayed in red,
illustrating, how the fracture mesh (green) cuts through the porous-medium matrix mesh (red).
The mesh width for this numerical experiment is set to $h=1/33$ in the porous-medium matrix and $h_{\gamma}=h_{\lambda}=1/200$ in the fractures.
In accordance with the previous numerical experiment, the element shapes are triangles in $\gamma$ and hexahedrons in $\Omega$. 
Fluid flow is modeled from left to right by non-homogeneous Dirichlet boundary conditions with values of 1 and 0, respectively.
The imposed boundary conditions on the remaining boundaries are homogeneous Neumann boundary conditions.\\
%%% RESULTS
The resultant dimensionless fluid pressure distribution in the fractures and the porous-medium matrix is depicted in Fig.~\ref{fig:randomDFN_150_fracs_results3D}. 
Additionally, three observation planes A, B, and C are defined to observe the fluid pressure in each plane (Fig.~\ref{fig:randomDFN_150_fracs_results2D}). 
Although the fluid pressure field is clearly influenced by the fracture network, Planes~B and~C show that the porous-medium matrix still contributes considerably to the overall fluid flow through the entire fractured porous medium. 
These results demonstrate the presented method's capability of modeling fluid flow through complex fractured porous media.
%%%
%%%
%%% FIG  - RANDOM FRACTURE NETWORK MESHING
%%%
%%%
%%% FIG - RANDOM FRACTURE NETWORK RESULTS 3D
%%%
%%%
%%% FIG - RANDOM FRACTURE NETWORK RESULTS CONTOURS
%%%
%%%
%%%
%%% DISCUSSION AND CONCLUSION
%%%
%%%
\section{Conclusion} 
\label{sec:conclusion}
%%% OVERVIEW
This study builds on previous work to combine a Lagrange multiplier method with an $L^2$-projection variational transfer operator to numerically model single-phase fluid flow problems in 3D fractured porous media. 
The performance of the method is assessed by comparison with benchmark cases and numerical experiments in 2D and 3D. 
We also demonstrate the method's suitability for large-scale, realistic fracture-network realizations in 3D.\\
%%% BENCHMARK 2D
Our comparison with benchmark simulations shows good agreement with reference results from the literature. 
In general, the accuracy of non-conforming methods at fracture intersections is known to depend on the width of the porous-medium matrix mesh.
These findings are confirmed in this study by all of our numerical experiments, as we also observe small deviations in fluid pressures at fracture intersections for larger mesh widths.
However, the RMS error in the porous-medium matrix mesh decreases linearly with increasing mesh size, as desired. \\
%%% BENCHMARK 2D EXTRUDED
To study the presented method in 3D, the 2D benchmark case is extruded in the third direction, thereby enabling a comparison of 3D simulations with 2D benchmark results. 
Although the 3D case provides additional complexity for the transfer operator, results agree well with the benchmark results. 
Analogous to the 2D results, the 3D results also show the largest deviations from the reference results at fracture intersections. \\
%%% SMALL DFN
A 3D fracture network, containing seven fractures, is presented, demonstrating the applicability of circular fractures with random orientations, sizes, and fracture tips inside the model domain. 
These simulations are compared to reference results obtained with conforming mesh simulations. 
A convergence study refined the porous-medium matrix mesh while leaving the fracture mesh unaltered.  
The observed convergence rates show a linear behavior in the porous-medium matrix and the fracture domains, with slightly improved convergence for the highest mesh resolution. 
The largest errors occur at the location of the steepest fluid pressure gradients, for example, at the tips of fractures that are aligned parallel to the fluid pressure gradient. 
More specifically, the errors occur at the fracture tips farthest apart from one another and in the direction of the fluid pressure gradient. \\
%%% BIG DFN
Finally, a numerical experiment of a realistic discrete fracture network (DFM), with 150 fractures, demonstrates the capability to model single-phase flow through highly complex fractured porous media.
The obtained fluid pressure contours in the 3D domain show the expected behavior for a fractured porous medium, with high-permeability fractures, largely influencing the fluid flow through the numerical model domain. 
Despite a coarse porous-medium matrix mesh, the presented method achieves detailed representations of fractures embedded in a porous-medium matrix domain. \\
%%% SUMMARY 
Generally, the Lagrange Multiplier--L$^2$-projection method shows good agreement with benchmark cases and reference results, while yielding good convergence. 
In all cases, our results suggest that the matrix mesh of the porous-medium should be locally refined when accuracy is to be improved, particularly around fracture tips and at fracture intersections. \\
%%% OUTLOOK
Further research will focus on the development of mesh adaptivity algorithms that allow pressure-gradient- and fracture-location-dependent porous-medium matrix-mesh refinements. 
Additional extensions of the numerical approach presented here will target transient fluid flow computation for which the parallel-processing algorithm, employed in this study, can be optimized. 
%
%%%
%%% ACKNOWLEDGEMENT
%%%
\section*{Acknowledgment}
M.O.S., P.S., A.E., D.V., and S.B.R. thank the Werner Siemens Foundation for their endowment
of the Geothermal Energy and Geofluids group at the Institute of Geophysics, ETH Zurich. 
P.Z., M.G.C.N, and R.K. thank the SCCER-SoE program.
We gratefully acknowledge the discussion with Markus K\"{o}ppel regarding his implementation of the Lagrange multiplier method.
%
%%%
%%% CODE AVAILABILITY
%%%
\section*{Computer Code Availability}
All methods and routines, used for this study, are implemented with the open-source software library \emph{Utopia}~\cite{utopiagit}. \emph{Utopia}'s lead developer is co-author Patrick~Zulian at USI~Lugano, Switzerland. 
Co-developers are Alena Kopani{\v c}{\'a}kov{\'a}, Maria Chiara Giuseppina Nestola, Andreas Fink, Nur Fadel, Victor Magri, Teseo Schneider, and Eric Botter. \\
The contact address and e-mail of Patrick Zulian are as follows:
\begin{itemize}
\item[] Institute of Computational Science\\
Universit{\`a} della Svizzera italiana (USI - University of Lugano)\\
Via Giuseppe Buffi 13\\
CH-6904 Lugano
\item[] patrick.zulian@usi.ch
\end{itemize}
\emph{Utopia} was first available in 2016, the programming language is \emph{C++} and it can be accessed through a git repository or a docker container on:
\begin{itemize}
\item[] https://bitbucket.org/zulianp/utopia (18.6~MB),
\item[] https://hub.docker.com/r/utopiadev/utopia.
\end{itemize}
The software dependencies are as follows:
\begin{itemize}
\item[] PETSc (https://www.mcs.anl.gov/petsc/),\\ must be compiled with \emph{MUMPS} enabled
\item[] libMesh for the FE module (https://github.com/libMesh)
\end{itemize}
There are no hardware requirements given by \emph{Utopia}.
Potential hardware or software requirements of the underlying libraries \emph{libMesh} and \emph{PETSc} are not stated here. 
%
%%%
%%% ORCID
%%%
\section*{ORCID}
P. Sch\"{a}dle - https://orcid.org/0000-0002-5485-2028\\
P. Zulian - https://orcid.org/0000-0002-5822-3288\\
D. Vogler - https://orcid.org/0000-0002-0974-9240\\
S. Bhopalam R. - https://orcid.org/0000-0001-7221-695X\\
M.G.C. Nestola - https://orcid.org/0000-0002-5700-0306\\
A. Ebigbo https://orcid.org/0000-0003-3972-3786\\
R.H. Krause - https://orcid.org/0000-0001-5408-5271\\
M.O. Saar - https://orcid.org/0000-0002-4869-6452\\
%
%%%
%%% BIBLIOGRAPHY
%%%
\bibliographystyle{abbrvnat}
\bibliography{bibliography.bib}
\end{document}